\input ./style/arxiv-general.cfg
\documentclass[aoas,MSNbibl,nameyear,seceqn,dvips]{arximspdf}
\makeatletter
   \@ifpackageloaded{graphicx}{}{\usepackage{graphicx}}
\makeatother
\usepackage{dcolumn}

\doi{10.1214/15-AOAS846}
\volume{9}
\issue{3}
\pubyear{2015}
\firstpage{1601}
\lastpage{1620}
\docsubty{FLA}

\makeatletter
\newcolumntype{d}[1]{D{.}{.}{#1}}
\newcommand{\bxij}{\mathbf{x}_{ij}}
\newcommand{\bxijo}{\mathbf{x}_{i,j+1}}
\newcommand{\byij}{\mathbf{y}_{ij}}
\newcommand{\bS}{\mathbf{S}}
\newcommand{\bfv}{\mathbf{f}}
\newcommand{\bZ}{\mathbf{Z}}
\newcommand{\bo}{\mathbf{0}}
\newcommand{\bI}{\mathbf{I}}
\newcommand{\bY}{\mathbf{Y}}
\newcommand{\bX}{\mathbf{X}}
\newcommand{\btheta}{\bolds{\theta}_{xi}}
\newcommand{\bthetaa}{\bolds{\theta}}
\newcommand{\bmu}{\bolds{\mu}}
\newcommand{\bSig}{\bolds{\Sigma}}
\newcommand{\bSige}{\bolds{\Sigma}_\varepsilon}

\newcommand{\bLamb}{\bolds{\Lambda}}
\newcommand{\bepsj}{\bolds{\varepsilon}_{ij}}
\newcommand{\bsh}{\setminus}
\newcommand{\bsigma}{\bolds{\Sigma}}
\newcommand{\bb}{\mathbf{b}}
\newcommand{\bw}{\mathbf{w}}
\newcommand{\bT}{\mathbf{T}}

\newcommand{\bbeta}{\bolds{\beta}}
\makeatother

\begin{document}
\begin{frontmatter}

\title{Bayesian analysis of ambulatory blood pressure dynamics
with application to irregularly spaced~sparse~data}
\runtitle{Bayesian analysis of ambulatory blood pressure dynamics}

\begin{aug}
\author[A]{\fnms{Zhao-Hua}~\snm{Lu}\thanksref{m1,T1}\ead[label=e1]{zhaohua.lu@gmail.com}},
\author[B]{\fnms{Sy-Miin}~\snm{Chow}\thanksref{m2,T2}\ead[label=e2]{symiin@psu.edu}\ead[label=u1,url]{http://www.personal.psu.edu/quc16/}},
\author[C]{\fnms{Andrew}~\snm{Sherwood}\thanksref{m3}\ead[label=e3]{andrew.sherwood@duke.edu}}\\
\and
\author[A]{\fnms{Hongtu}~\snm{Zhu}\corref{}\thanksref{m1,T3}\ead[label=e4]{htzhu@email.unc.edu}\ead[label=u2,url]{http://www.bios.unc.edu/research/bias/people.html}}
\runauthor{Lu, Chow, Sherwood and Zhu}
\thankstext{T1}{Supported by NSF Grant BCS-0826844.}
\thankstext{T2}{Supported by NSF Grants SES-1357666 and BCS-0826844, NIH Grant R01GM105004,
Penn State Quantitative Social Sciences Initiative and UL TR000127 from the National Center for Advancing Translational Sciences.}
\thankstext{T3}{Supported by NIH Grants RR025747-01, MH086633 and 1UL1TR001111, and NSF Grants SES-1357666, DMS-14-07655 and  BCS-0826844.}
\affiliation{University of North Carolina at Chapel Hill\thanksmark
{m1}, Pennsylvania State University\thanksmark{m2} and Duke
University\thanksmark{m3}}
\address[A]{Z.-H. Lu\\
H. Zhu\\
Department of Biostatistics\\
University of North Carolina at Chapel Hill\\
Chapel Hill, North Carolina 27599\\
USA\\
\printead{e1}\\
\phantom{E-mail:\ }\printead*{e4}\\
\printead{u2}}

\address[B]{S.-M. Chow\\
Department of Human Development\\
\quad and Family Studies\\
Pennsylvania State University\\
118 Henderson South Building\\
University Park, Pennsylvania 16803 \\
USA\\
\printead{e2}}

\address[C]{A. Sherwood\\
Department of Psychology and Neuroscience\\
Duke University\\
4569 Hosp South\\
Campus Box 3119 Med Ctr\\
Durham, North Carolina 27708\\
USA\\
\printead{e3}}
\end{aug}

%
\received{\smonth{6} \syear{2014}}
%
\revised{\smonth{2} \syear{2015}}

%
\begin{abstract}
Ambulatory cardiovascular (CV) measurements provide valuable insights
into individuals' health conditions in ``real-life,'' everyday
settings. Current methods of modeling ambulatory CV data do not
consider the dynamic characteristics of the full data set and their
relationships with covariates such as caffeine use and stress. We
propose a stochastic differential equation (SDE) in the form of a dual
nonlinear Ornstein--Uhlenbeck (OU) model with person-specific covariates
to capture the morning surge and nighttime dipping dynamics of
ambulatory CV data. To circumvent the data analytic constraint that
empirical measurements are typically collected at irregular and much
larger time intervals than those evaluated in simulation studies of
SDEs, we adopt a Bayesian approach with a regularized Brownian Bridge
sampler (RBBS) and an efficient multiresolution (MR) algorithm to fit
the proposed SDE. The MR algorithm can produce more efficient MCMC
samples that is crucial for valid parameter estimation and inference.
Using this model and algorithm to data from the Duke Behavioral
Investigation of Hypertension Study, results indicate that age,
caffeine intake, gender and race have effects on distinct dynamic
characteristics of the participants' CV trajectories.
\end{abstract}

%
\end{frontmatter}
\setcounter{footnote}{3}
\section{Introduction}\label{sec1}



Coronary heart disease (CHD) is the leading cause of morbidity and
mortality in older adults, and instances of deaths due to CHD and
stroke are estimated by the Centers for Disease Control and Prevention
(CDC) as ``nearly twice the number of lives claimed by cancer or
collectively by World War II, and the Korean and Vietnam conflicts''
[\citet{CDC99a}]. There has been increasing evidence that cardiovascular
(CV) measures, such as ambulatory blood pressure (ABP) taken in
everyday, nonlaboratory settings, provide better diagnostic and
prognostic value than multiple clinic blood pressure (BP) readings
[\citet{doi:10.1056/NEJMoa022273,enlighten18502}], and are indicative of
the occurrence of multiple CV events [\citet
{Beckham09a,Willich199265,Muller01041989}].



ABP and related CV activities (CA) have well-established circadian
patterns, characterized by rises in early morning, culminating in a
plateau around noon, and followed by nocturnal (nighttime) dipping.
While nighttime BP has been found to be a stronger predictor of
cardiovascular risk than clinic or daytime ABP [\citet{Hansen01012011}],
increasing evidence has pointed to the importance of also considering
morning surges in ABP in addition to nighttime BP [\citet
{Kario03a,Verdecchia12a}]. The importance of studying the dynamics of
ABP is further reflected in the inclusion of trend reports in popular
ABP measurement tools such as the dabl system [\citet{JOIM:JOIM2356}],
which provide indices such as time-weighted measures of variability,
measures of nocturnal dip, morning surge, peak as well as trough
levels, and smoothness of BP curves, among many other indices of CV
events [\citet{Dolan01032006,Rothwell2010895}]. Despite the richness of
the dynamic information in ABP data, diagnosis/prognosis involving ABP
is typically performed on levels of ABP obtained from isolated segments
of the data. As an example, morning surge is typically defined as a
rise in BP $>$ 55~mmHg from the lowest nighttime reading [for a review
see \citet{JOIM:JOIM2356}]. In a similar vein, individuals are
identified as exhibiting BP nondipping---a commonly used prognostic
indicator of CV morbidity and mortality for both hypertensive and
nonhypertensive individuals---when they show $<$10\% fall in systolic
BP (SBP) from day to night [\citet
{doi:10.1001/jama.295.24.2859,Fagard01012008}]. Such conventional
approaches of analyzing ABP rely solely on levels of BP during selected
time windows, and
utilize levels of BP at a single time point (e.g., the lowest nighttime
reading), which are less than ideal given the noisy nature of BP and
other CV measures. In addition, some of the more subtle individual
differences in dynamic characteristics of CV measures, such as the
surge and dipping rates of CV measures, are completely bypassed.


One possible way to extract more dynamic information from individuals'
full time series of CV measures is to analyze such data in the context
of a stochastic differential equation (SDE) model. The SDE of choice
has to capture critical aspects of CV dynamics while providing a
platform to relate these dynamic attributes to individual difference
characteristics such as stress levels, age and so on. %
To enable SDE modeling of multiple measures of population CV activities
(e.g., systolic BP, diastolic BP and heart rate), we propose a latent
SDE in the form of a dual nonlinear Orstein--Uhlenbeck (OU) model with
person-specific dynamic effects. This modeling framework provides a
direct way to (i) represent the unobserved dynamics of CV activities
based on noisy multivariate measurements from multiple subjects; (ii)
accommodate subject-specific, irregularly spaced discrete time points,
particularly the sparse measurements at night to minimize disruptions
to the participants' sleep schedules; and (iii) allow the evaluation of
questions pertaining to the dynamics of ambulatory CV data, including
individual differences in morning surge and nighttime dipping patterns.

Estimation and inference of SDE models using ambulatory CV data are
challenging. Due to the intractability of the proposed SDE, we employ
discretization approximation [\citet{Pedersen1995}].
Unfortunately, real-life ambulatory CV data are characterized by much
sparser and irregularly spaced time intervals than those investigated
in most simulation studies involving nonlinear SDE models [\citet
{kou2011,Lindstrom2012}]. Achieving reasonable estimation properties
necessitates the use of a large number of imputations between
subsequent observed intervals, a procedure that quickly becomes
inefficient for the kind of data considered. We develop an efficient
regularized Brownian bridge sampler (RBBS) and multiresolution (MR)
algorithm to fit the proposed SDE model.

\section{Data analytic and methodological issues}

\begin{figure}[t]

\includegraphics{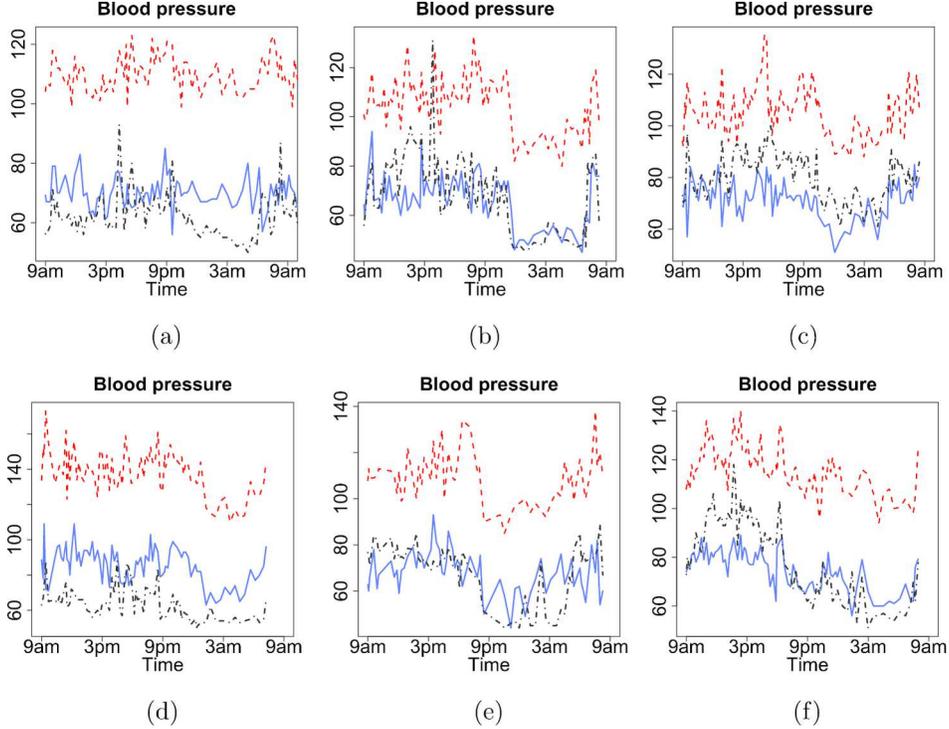}

\caption{Trajectories of SBP, DBP and heart rate of six subjects in the
case study, which are the dashed, solid and dot-dashed curves, respectively.}
\label{rti-fig1}
\end{figure}

The empirical data in our study consist of CV measures from multiple
subjects. Figure~\ref{rti-fig1} shows the data from six subjects from
the study. The dashed, solid and dot-dashed curves are SBP, DBP and
heart rate, respectively.
All three measures are characterized by relatively systematic circadian
rhythms and some subject-specific characteristics.

%
%
%
%
%

Over a 24-hour period, all three measures typically decrease to their
lowest points during nighttime sleep and increase rapidly upon rising
in the morning. However, the circadian patterns and magnitudes of
change, which are related to cardiovascular risk, show considerable
between-subject heterogeneities. Consider the baseline levels around
which the trajectories fluctuate at daytime and nighttime as two
equilibria. First, the differences between two equilibria may be
different for different subjects. For instance, the subject in
Figure~\ref{rti-fig1}(a) shows less difference in his/her daytime and nocturnal
equilibria than the subject depicted in Figure \ref{rti-fig1}(b), thus signaling
less dipping (or poorer recovery). Second, even if the differences
between two equilibria are similar, the magnitudes of the equilibria
can be different [e.g., Figure~\ref{rti-fig1}(c) and (d)].
Such cases demonstrate that using the differences between the
equilibria alone to analyze BP data may obscure important dynamic
features of the data. Third, the rates of change during dipping and
surge may be different and also show various degrees of asymmetry
across subjects. Both the morning surge rate and nocturnal dipping rate
in Figure~\ref{rti-fig1}(b) are\vadjust{\goodbreak} large; the nocturnal equilibrium, in
particular, is attained very quickly and efficiently. In comparison,
the subject in Figure~\ref{rti-fig1}(e) shows quick dipping and slower
surge than the subject in Figure~\ref{rti-fig1}(b), while Figure~\ref
{rti-fig1}(f) shows the reverse change patterns. Last, SBP, DBP and HR
often share common features/circadian trends within subjects, thus
motivating us to use a latent process to characterize their common dynamics.

\begin{figure}[b]

\includegraphics{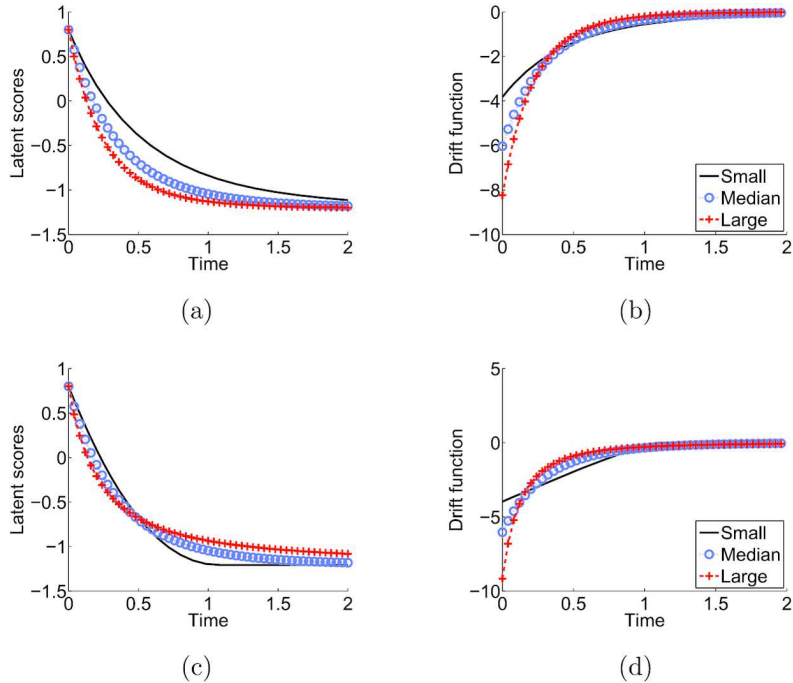}

\caption{The latent scores and corresponding drift functions given
different parameter values in~(\protect\ref{dwm}). The first row of plots
features the expected trajectories and the drift function with
different relative change rates, while the second row is those with
different change instabilities. The cross, circle and solid curves are
generated from large, medium and small relative change rates or change
instabilities, respectively.}\vspace*{-3pt}
\label{rti-fig}
\end{figure}

We formulate a Latent Stochastic Differential Equation Model (LSDEM) to
capture subject-specific (i.e., covariate-dependent): (i) daytime and
nighttime CV equilibria, (ii) nighttime dipping rate and morning surge
rate, and (iii) dipping and surge patterns. The Ornstein--Uhlenbeck (OU)
process is widely used to model a stochastic process that fluctuates
around an equilibrium [\citet
{uhlenbeck1930theory,ricciardi1979ornstein,EVO:EVO1619}].
The CV patterns observed in Figure~\ref{rti-fig1} motivated us to
employ a modified dual-OU process model expressed as
%
\begin{eqnarray}\label{dwm}\quad
dx_i(t)&=&\bigl[I(t\in M_i)\beta_{i1}^*\bigl(
\beta_{i2}^* - x_i(t)\bigr)^{\beta
_{i3}^*}+I(t\notin
M_i)\beta_{i4}^*\bigl(\beta_{i5}^* -
x_i(t)\bigr)^{\beta
_{i6}^*}\bigr]\,dt
\nonumber
\\[-8pt]
\\[-8pt]
\nonumber
&&{}+\sqrt{\psi} \,dB_t,
\end{eqnarray}
where $x_i(t)$ is the $i$th subject's latent CV activity at time $t$,
$M_i$ is subject $i$'s daytime period, determined based on commonly
used time windows and subject-specific data cues (which include, e.g.,
the preawakening window during which participants have not started
engaging in their everyday routines but may have started to show rises
in CV activities), and $\psi$ is the variance of the Wiener process,
commonly referred to as the diffusion parameter. The $\beta_{ij}^* =
\bw
_i^T\bbeta_j$ for
$j = 1,\ldots,6$ are subject-specific dynamic parameters, where $\bw_i$
and $\bbeta_j$ are vectors of covariates and slope parameters,
respectively. Moreover, $\beta_{i1}^*$, $\beta_{i3}^*$, $\beta_{i4}^*$
and $\beta_{i6}^*$ are assumed to be positive.

The conventional OU process is a special case of (\ref{dwm}) with
$\beta
_{i1}^*=\beta_{i4}^*$, $\beta_{i2}^*=\beta_{i5}^*$, and $\beta
_{i3}^*=\beta_{i6}^*=1$. The drift function in (\ref{dwm}) quantifies
the absolute rates of change. The
$\beta_{i2}^*$ and $\beta_{i5}^*$ are used to represent the daytime and
nighttime equilibria, respectively, around which the CV trajectories
fluctuate. It is assumed that the rate of change of CV activities is
proportional to the difference between the current CV activities and
the equilibrium, and the relative rates of change in daytime and
nighttime are modeled by $\beta_{i1}^*$ and $\beta_{i4}^*$,
respectively. The first row of plots in Figure~\ref{rti-fig}
illustrates the effect of $\beta_{i4}^*$\vadjust{\goodbreak} during dipping. The left plots
show the trajectories simulated from (\ref{dwm}) with 3 levels of
$\beta
_{i4}^*$ and $\psi=0$. The right plots show the corresponding values of
the drift function, where a value of zero on the ordinate represents no
change in the value of $x_i(t)$ for a specific value of $dt$. The
larger $\beta_{i4}^*$ is, the faster CV activities go to the night
equilibrium. The $\beta_{i3}^*$ and $\beta_{i6}^*$ affect the shape of
the drift function. We interpret them as the change instability parameters.
The second row of Figure~\ref{rti-fig} illustrates the effect of
$\beta
_{i6}^*$ during dipping. Larger $\beta_{i6}^*$ leads to quicker dipping
at the beginning. However, the rate of change decreases more quickly as
CV activities approach the equilibrium. Hence, the rate of change is
less stable. In comparison, the rate of change decreases more slowly
for smaller $\beta_{i6}^*$ values. The interpretations of $\beta
_{i1}^*$ and $\beta_{i3}^*$ are similar for morning surge.

We assume the diffusion parameter, $\psi$, to be constant in (\ref
{dwm}). Substantively, it is reasonable to assume that the main
variations of CV activities among subjects stem from differences in
mean levels of CV activities during daytime and nighttime, as well as
the transitions in between, which are mainly characterized by the drift
function in (\ref{dwm}). Consequently, the diffusion function only
characterizes the fluctuation around the equilibriums of CV activities,
the scale of which is comparatively small and the differences across
subjects are comparatively insignificant. Thus, we only consider a
constant diffusion function in the current analysis.



Methodological challenges associated with fitting SDE models such as
that shown in (\ref{dwm}) can become formidable in the presence of
sparse and wide-ranging time intervals. Among methods for estimating
parameters in SDEs [for a review see \citet{sorensen2004parametric}],
likelihood-based methods have received much attention, but they require
solutions of transition density functions of SDEs that are analytically
available for only a very limited class of SDEs.
Methods to circumvent this difficulty include closed-form expansion of
the transition density [\citet{ait2008closed}], exact simulation method
[\citet{beskos2006exact,sermaidis2011markov}] and discrete
Euler--Maruyama approximation with data augmentation between observed
time points [\citet{Pedersen1995,durham2002,zhu2011jasa}]. %
The last approach is popular due to its general applicability, but the
time intervals in studies involving ambulatory measures are usually too
large to enable accurate estimation. Specifically, to achieve
reasonable approximation accuracy, the time points between successive
observations need to be augmented with missing data [\citet
{Elerian2001}]. Increasing the number of the augmented time points not
only leads to better approximation, but also increases the dependency
among modeling parameters and the diffusion paths [\citet{Elerian2001}],
leading to slower convergence of the data augmentation algorithms for
estimation and inference. In situations involving irregularly spaced
time points, this problem is exacerbated because large time intervals
require more imputed time points to insure approximation accuracy.
Block updating algorithms have been proposed to alleviate the
dependency among missing data at augmented time points, but the
dependency between parameters in the diffusion function and the
diffusion paths remains problematic in most SDEs [\citet
{Roberts01102001}]. To handle these problems, we adopt a Bayesian
approach for parameter estimation and inference, and utilize two
efficient MCMC algorithms, namely, block updating with regularized
Brownian bridge sampler (RBBS) and a multiresolution (MR) algorithm,
derived and adapted respectively from \citet{Lindstrom2012} and \citet
{kou2011}, to fit the proposed SDE model.

Bayesian approaches have served as promising tools for the estimation
and inference for SDEs [\citet
{Elerian2001,Roberts01102001,durham2002,Golightly20081674}]. Recently,
\citet{stramer2011bayesian} proposed the use of two different
simula\-tion-based approximations to achieve better approximation of the
likelihood with fewer imputations. \citet{golightly2011bayesian}
developed particle MCMC algorithms to update the processes globally and
sequentially, avoiding the dependency problem as the processes and
parameters are sampled jointly.

\section{Latent stochastic differential equation models (LSDEMs)}
While the SDE model shown in (\ref{dwm}) is designed specifically to
capture ambulatory CV dynamics, our estimation algorithms are
applicable to a broader class of models that includes other linear and
nonlinear latent stochastic models (LSDEMs) as special cases. The
general model is a hierarchical model consisting of two parts: (i) a
factor analysis model that relates a vector of latent variables to
their noisy, observed counterparts, and (ii) a SDE model for describing
the changes in the latent variables.
\subsection{Factor analysis model}
Let $\mathbf{x}_{i}(t)$ be a $q \times1$ vector of latent processes
of interest,
where the indices $i$ and $t$, respectively, denote individuals and
time; $\mathbf{y}_{i}(t)$ is a $p \times1$ vector of observed
processes (e.g.,
SBP, DBP and heart rate in our study). The latent variables in $\mathbf
{x}_{i}(t)$
are measured indirectly through $\mathbf{y}_{i}(t)$ based on the
measurement model:
%
\begin{equation}
\label{me0} \mathbf{y}_{i}(t)=\bmu+\bLamb\mathbf{x}_{i}(t)+
\bolds{\varepsilon}_{i}(t),
\end{equation}
where $\bmu$ is a $p\times1$ vector of intercepts, $\bLamb$ is a
$p\times q$ loading matrix, and $\bolds{\varepsilon}_{i}(t)$ denotes a
$p\times1$ vector of
measurement error processes that is independent of $\mathbf{x}_{i}(t)$.
In most human dynamics studies, however, we only measure $\mathbf
{y}_{i}(t)$ at
irregularly spaced time points $t_{ij}$ for $j=1, \ldots, T_i$ and
$i=1, \ldots, n$,
where $t_{ij}$ is the $j$th time point for the $i$th individual.
The measurement model (\ref{me0})
at $t_{ij}$ is given by 
$\byij=\bmu+\bLamb\bxij+\bepsj$, $1\leq j \leq T_i$, $1\leq i \leq n$,
where
$\bxij=\mathbf{x}_i(t_{ij})$, $\bepsj={\boldmath\varepsilon}_{i}(t_{ij})$
and $\mathbf{y}_{ij}=\mathbf{y}_i(t_{ij})$. The $\bepsj$ is
independent of
$\bxij$ and follows $N(0,\bSige)$, in which $\bSige$ is a diagonal
matrix with diagonal elements $(\sigma^2_{\varepsilon1},\ldots,\sigma
^2_{\varepsilon p})$.

\subsection{SDE model for latent change processes}
We consider SDEs for delineating the dynamics of latent variables. Let
$d$ be the differential operator. The SDE model of interest is given by
%
\begin{equation}
\label{SDE1} d\mathbf{x}_{i}(t)= \bfv\bigl(\mathbf{x}_{i}(t),
\btheta\bigr)\,dt+\bS\bigl(\mathbf {x}_{i}(t),\btheta\bigr)\,d
\mathbf{B}_{i}(t),
\end{equation}
where $\bfv(\cdot)=(f_1(\cdot),\ldots,f_q(\cdot))$ is a $q \times1$
vector of drift functions, $\bS$ is a $q \times q$ matrix of diffusion
functions, and $\mathbf{B}_{i}(t)$ is a $q \times1$ vector of
standard Wiener
processes, whose increments, d$\mathbf{B}_{i}(t)$, are Gaussian
distributed with
zero means and variances that increase with the length of time
interval, $dt$. Moreover, $\btheta=g(\bw_i,\bb)$, where $g(\cdot)$
is a
known function with a vector of parameters $\bb$ and covariates $\bw
_i$. One important question of our study is to identify predictors that
can explain the heterogeneities in dynamics across subjects. The
$\btheta$ is used to characterize subject-specific differences in
change as related to known, person-specific covariates. A~heuristic
interpretation of $\bfv$ and $\bS$ is that $\bfv$ governs the local
changes (i.e., \textit{drift} rates) in $\mathbf{x}_{i}(t)$ over
$dt$, whereas
$\bS
$ governs the variance of local changes, or in other words, the \textit
{diffusion} rates.

Since most SDEs in (\ref{SDE1}) do not have analytical solutions, it is
common to employ a discretized approximation, such as Euler--Maruyama, %
at selected time points to form an approximate likelihood for model
(\ref{SDE1}):
%
\begin{equation}
\Delta\bxij= \bfv(\bxij,\btheta)\Delta t_{ij}+{\Delta
t_{ij}^{1/2}}\bS (\bxij,\btheta)\bZ_{ij}
\label{dynfun}
\end{equation}
for $0\leq j < T_i$ and $1\leq i \leq n$, where $\Delta\bxij=\bxijo
-\bxij$, $\Delta t_{ij}=t_{i,j+1}-t_{ij}$, and $\bZ_{ij}$ follows a
multivariate Gaussian distribution $N(\bo,\bI_q)$, in which $\bI_q$ is
a $q\times q$ identity matrix. When $j=0$, the initial observations of
the processes $\mathbf{x}_{i0}$ are assumed to be known for all $i$.

Empirical data are usually sampled at relatively sparse intervals, so
the Euler--Maruyama approximation (\ref{dynfun}) performed only at
empirically observed time points usually leads to poor likelihood
approximation [\citet{Elerian2001}].
To increase the accuracy of the approximation (\ref{dynfun}), we
impute $\mathbf{x}_{i}(t)$ at additional unobserved time points
between $t_{ij}$'s
as missing data.
In practice,
the number of imputed missing data between two observed time points
determines the \textit{resolution} and accuracy of the approximation
(\ref{dynfun}). Let $\bX^{(0)}=(\bX^{(0)}_1,\ldots,\bX^{(0)}_n)$ and
$\bX_{i}^{(0)}=(\mathbf{x}_{i1},\ldots,\mathbf{x}_{iT_i})$ be the processes
at the
observed time points for the $i$th individual.
More time points are imputed between two adjacent time points with
larger $\Delta t_{ij}$. The time intervals after imputation are close
to the minimal time interval before imputation. Denote $t_{ij}^{(1)}$
to be the $i$th subject's $j$th time point after imputation,\vspace*{1pt} $\Delta
t_{ij}^{(1)}=t_{i,j+1}^{(1)}-t_{ij}^{(1)}$, and
$x_{ij}^{(1)}=x_i(t_{ij}^{(1)})$ for $0\leq j < T_i^{(1)}$ and $1\leq i
\leq n$.
Let this imputation be the $1$st resolution. The accuracy of the
Euler--Maruyama approximation can be refined by increasing the number of
imputed time points, that is, increasing the resolution. The $k$th
resolution is constructed by imputing one time point between two
adjacent time points at the $(k-1)$th resolution. For notational
simplicity, we assume that only 1 time point is imputed between each
pair of adjacent observed time points to construct the $1$st
resolution. However, the algorithm is applicable to general situations
with heterogeneous imputation at the $1$st resolution.
Let $k^*=2^{k}$. At the $k$th resolution, let $\mathbf
{x}^{(k)}_{ij}=\mathbf{x}
_i(t_{ij}^{(k)})$, and
\begin{eqnarray}
t_{ij}^{(k)}=t_{is}^{(k-1)}+
\bigl(t_{i,s+1}^{(k-1)}-t_{is}^{(k-1)}\bigr)
\frac
{j-2s}{2}
\nonumber\\
\eqntext{\mbox{for } 2s\leq j\leq2(s+1)\mbox{ and } s=0,
\ldots,T_i^{(k-1)}.}
\end{eqnarray}
%
Consequently, $\mathbf{x}^{(k)}_{i,sk^*}=\mathbf{x}_{is}$. Let
$\Delta
t_{ij}^{(k)}=t_{i,j+1}^{(k)}-t_{ij}^{(k)}$, $\bX_{i}^{(k)}=(\mathbf{x}^{(k)}
_{i1},\ldots
,\mathbf{x}^{(k)}_{i,T_i^{(k)}})$, and $\bX^{(k)}=(\bX
^{(k)}_1,\ldots,\bX^{(k)}_n)$.
The approximated transition density is 
%
\begin{equation}
\label{dpconddist} P_k\bigl(\bxijo^{(k)}|\bxij^{(k)},
\btheta\bigr)= \phi_q \bigl(\bxij ^{(k)}+\bfv \bigl(
\bxij^{(k)} ,\btheta\bigr)\Delta t_{ij}^{(k)},\Delta
t_{ij}^{(k)}\bSig\bigl(\bxij^{(k)} ,\btheta\bigr)
\bigr),
\end{equation}
where $\bSig(\bxij^{(k)},\btheta)=\bS(\bxij^{(k)},\btheta)\bS
(\bxij
^{(k)},\btheta
)^T$ and $\phi_q(\bmu, \bSig)$ denotes the density of a $q$-dimensional
Gaussian random vector with
mean vector $\bmu$ and covariance matrix $\bSig$.

\section{Bayesian estimation and inference with MCMC algorithms}
Let $\bY_i=(\mathbf{y}_{i1},\ldots,\mathbf{y}_{iT_i})$, $\bY=(\bY
_1,\ldots,\bY_n)$
and $\bthetaa=\{\bb,\bmu,\bLamb,\bSige\}$. We augment $\bX^{(k)}$
to the
observed data $\bY$, and then use MCMC algorithms [\citet
{Hastings1970,Geman1984}] to sample $P_k(\bthetaa,\bX^{(k)}|\bY
)\propto
P_k(\bY,\bX^{(k)}|\bthetaa)P(\bthetaa)$. Generally, any distributions
representing the prior information could be used.
We assume that $P(\bthetaa)=P(\bmu)P(\bLamb,\bSige)P(\bb)$, and
use the
prior distributions leading to standard full conditional distributions,
%
\begin{eqnarray}\label{mesprior}
P(\mu_r)&=&\phi_1\bigl(\mu_{r0},
\sigma^2_{\mu0}\bigr),\qquad P(\bLamb_r)=\phi
_p\bigl(\bLamb _{0r},\sigma^2_{\varepsilon r}
\bSig_{\Lambda r}\bigr),
\nonumber
\\[-8pt]
\\[-8pt]
\nonumber
  P\bigl(\sigma ^{2}_{\varepsilon
r}
\bigr)&=&\operatorname{IG}(a_{1r},a_{2r}),
\end{eqnarray}
where $r=1,\ldots,p$, $\mu_r$ is the $r$th row of $\bmu$, and
$\bLamb
_r^T$ is the
$r$th row of $\bLamb$. The $\mu_{r0}$, $\sigma^2_{\mu0}$, $\bLamb
_{0r}$, $a_{1r}$, $a_{2r}$ and positive definite matrix $\bSig
_{\Lambda
r}$ are hyperparameters, the values of which are assumed to be given by
prior information. The $\operatorname{IG}(\cdot,\cdot)$ stands for the inverse gamma
distribution.
The $\mu_r$ estimate is more robust with different signal strengths,
for example, scale of $\sigma^{2}_{\varepsilon r}$. Hence, its prior
distribution is assumed to be independent of $\sigma^{2}_{\varepsilon r}$.
As $\bb$ includes parameters that are involved in the functions $\bfv
(\cdots)$ and $\bS(\cdots)$, the corresponding prior distributions have
to be tailored specifically to the dynamic model considered.

For sparsely spaced data, efficient sampling of $\bX^{(k)}$ is very
challenging. Most approaches based on the Euler--Maruyama approximation
only use one resolution $k$. Larger $k$ results in better
approximation, but it increases computational costs from two aspects.
First, the dimension of $\bX^{(k)}$ increases with $k$. Second, the
MCMC efficiency decreases dramatically because smaller $\Delta
t_{ij}^{(k)1/2}\bS(\bxij^{(k)},\btheta)$ leads to high correlations
among $\bxij^{(k)}$. Consequently, more iterations of the Gibbs sampler
are required to obtain ``good'' MCMC samples that cover the entire
parameter/unobserved components space. Choosing $k$ to strike an
effective balance between approximation accuracy and sampling
efficiency is challenging, especially for nonlinear processes, where a
large $k$ is usually required.
We develop an efficient multiresolution MCMC algorithm [\citet
{kou2011}] to address such challenging issues.

\subsection{Multiresolution (MR) algorithm}\label{mrsec}

The MR algorithm [\citet{kou2011}] provides one way to circumvent the
inadequacies of using one specific resolution scheme by consolidating
samples obtained at multiple resolutions. They proposed the MR
algorithm for stochastic
processes observed at discrete time points for a single subject. We
will extend the MR algorithm for latent processes for population data.

The MR approach is a mixture of a series of local samplers and a global
sampler, and generates samples for every resolution sequentially. In
each iteration, the local samplers and the global sampler are chosen
with certain probabilities. The MR algorithm begins with the first
resolution with the least imputation. At each resolution, 
any MCMC algorithms designed for a single resolution
can be used as \textit{local samplers}, which explore the local
features of $P_k(\bthetaa,\bX^{(k)}|\bY)$. Starting from the second
resolution, a global sampler called ``cross-resolution sampler'' is
also used, which essentially performs an independent
Metropolis--Hastings (MH) update of $\bX^{(k)}$ and $\bthetaa$ jointly.
Let $\bT^{(k)}=\{t_{ij}^{(k)}|0\leq j \leq T_i^{(k)},1\leq i \leq n\}$ and
$\bX^{(k)\bsh(k-1)}$ be the processes at $\bT^{(k)}$ but not at $\bT
^{(k-1)}$.
The proposal distribution
$q(\bX^{(k)},\bthetaa)=q(\bX^{(k)\bsh(k-1)}|\bX
^{(k-1)},\bthetaa
)q(\bX
^{(k-1)},\bthetaa)$,
where $q(\bX^{(k-1)},\bthetaa)=P_{k-1}(\bthetaa,\bX^{(k-1)}|\bY)$.
Practically, $(\bX^{(k-1)},\bthetaa)$ are empirically sampled from the
MCMC samples for $P_{k-1}(\bthetaa,\bX^{(k-1)}|\bY)$.
The proposal samples are weighted in order that the target distribution
follows $P_{k}(\bthetaa,\bX^{(k)}|\bY)$. The cross-resolution sampler
is independent of the current state of $\bX^{(k)}$ and $\bthetaa$ and
overcomes the degeneracy caused by increasing dependency among $\mathbf{x}
_{ij}^{(k)}$ as $k$ increases.
Moreover, the empirical samples from a coarser resolution
$P_{k-1}(\bthetaa,\bX^{(k-1)}|\bY)$ have lower autocorrelation.
Hence, a cross-resolution sampler could move across the space of $\bX
^{(k)}
$ and $\bthetaa$ faster.
It is worth noting that even though the cross-resolution sampler for
$(\bX^{(k)},\bthetaa)$ is based on the MCMC samples of $\bX^{(k-1)}$,
the MCMC samples of $\bX^{(k)}$ at $\bT^{(k-1)}$ are partially
different from those of $\bX^{(k-1)}$ because local samplers are also
used with nonzero probability.
More details of the MR algorithm and cross-resolution sampler can be
found in S1.2 and S1.3 in supplementary material [\citet{supp}], and
in \citet{kou2011}.

\subsubsection{Local updating algorithms}\label{LUA}

Let $\tilde{k}=k^*-1$. We use two local samplers to generate posterior
samples of $\bX^{(k)}$ including (i) a 1-step RBBS that samples
$\mathbf{x}
^{(k)}_{ij}$ at each time point; and (ii) a block updating scheme for
$\bX_{is}=(\mathbf{x}_{i,sk^*},\ldots,\mathbf{x}_{i,(s+1)k^*})$
based on a
$(\tilde
{k} +2)$-step RBBS. Related samplers were proposed for univariate and
multivariate nonlinear SDEs for a single subject with observed
processes, respectively, in \citet{kou2011} and \citet{Lindstrom2012}.
We extend these samplers to handle latent SDEs for population data. The
general idea of the RBBSs is to construct a multivariate normal
proposal distribution for $\mathbf{x}^{(k)}_{ij}$ and $\bX^{(k)}_{is}$
sequentially, and use the MH algorithm. \citet{Lindstrom2012} is an
extension of \citet{durham2002}'s results on nonlinear processes, in
which the drift functions dominate the diffusion functions. Users
should adjust the tuning parameter $\alpha$ according to specific
problems, for which \citet{Lindstrom2012} provided intuitive
suggestions. More information regarding these extensions is described
in~S1.4 in supplementary material [\citet{supp}]. 

\section{Case study}

We analyzed a set of 24-hour ambulatory CV data from the Duke
Biobehavioral Investigation of Hypertension study [\citet
{sherwoodnighttime2002}]. %
The data set consists of 179 men and women whose ages range from 25 to
45 years. Ambulatory BP and other CV measures were monitored using the
noninvasive AccuTracker II ABP Monitor (Suntech AccuTracker II,
Raleigh, NC) from around 9 AM until the same time in the following
morning. The monitors were programmed to measure four times an hour at
random intervals ranging from 12 to 28 minutes apart during waking
hours. During sleeping hours, the monitors were programmed to record
only two readings hourly, customized to fit the participants' sleep
habits. The study maintained participants' normal schedules and
documented a diary entry indicating posture, activity, location,
positive affect and negative affect at each reading. Mood states were
scored by circling a number on a 5-point Likert Scale, with 1
representing ``not at all'' and 5 representing ``very much.''

Covariates of interest used in $\mathbf{x}_i$ in model (\ref{dwm})
are (i)
mean caffeine consumption during daytime and nighttime; (ii) overall
negative emotion score calculated as the mean of ``Stress,'' ``Anger''
and ``Tense'' ratings throughout the entire day; (iii)~overall positive
emotion score calculated as the mean of ``Happy'' and ``In
control'' ratings throughout the day; (iv) gender; (v) race; and (vi)
age. In addition, it is assumed that the effects of caffeine
consumption during daytime (nighttime) only affect the CV activities in
the daytime (nighttime) through $\beta_{i1}^*$, $\beta_{i2}^*$ and
$\beta_{i3}^*$ ($\beta_{i4}^*$, $\beta_{i5}^*$ and $\beta_{i6}^*$).

Time was rescaled such that 1 unit represents 12 hours. The resulting
lengths of time intervals between two adjacent observed time points
range from $0.01$ to $0.48$, corresponding to a range of 0.72 minutes to
5.76 hours in time. To form the 1st resolution, imputed time points are
placed evenly between two observed time points. The time intervals
between the imputed time points are around $0.07$ (corresponding to 5 minutes).
To obtain the daytime and nighttime windows in (\ref{dwm}), we used
manual coding to extract subject-specific time windows for each
subject. Specifically, the daytime window for each subject was defined
to end when systematic dipping of BP and heart rate were observed,
while the nighttime window was defined to end when systematic rise in
BP and heart rate were observed.

The measurement model for the latent process $x_{ij}$ at $j=sk^*$ is
given by
%
\begin{equation}
\label{dwmm} %
\pmatrix{ y_{is1}, y_{is2},
y_{is3}}
^T= %
\pmatrix{ 1, \lambda_1, \lambda_2 }
^T x_{ij} + %
\pmatrix{
\varepsilon_{is1}, \varepsilon_{is2}, \varepsilon_{is3}
} %
^T,
\end{equation}
where $y_{is1}$, $y_{is2}$ and $y_{is3}$ are, respectively, SBP, DBP
and heart rate at the $s$th observed time for the $i$th subject. Each
$y_{isj}$, $j=1,2,3$, was standardized by the mean and standard
deviation calculated using all $i$ and $s$. Moreover, $(\varepsilon_{is1},
\varepsilon_{is2}, \varepsilon_{is3})^T\sim N[\bo,\operatorname{diag}(\sigma^2_{\varepsilon
1},\sigma^2_{\varepsilon2},\sigma^2_{\varepsilon3})]$, and the $1$ in the
loading matrix is fixed for identification.
The initial conditions of the latent SDEs were fixed to the estimated
factor scores of the subjects using the values of SBP, DBP and heart
rate at the first observed time point.
Without any prior information, vague prior distributions $\bbeta_j\sim
N(\bb_{0j},\bsigma_{0bj})I(S_j)$ and $\psi\sim \operatorname{IG}(a_{\psi1},a_{\psi
2})$ were assumed for the parameters in model (\ref{dwm}), 
where $\bb_{0j}=\bo$, $\bsigma_{0bj}=10^6\bI$, for $j=1,\ldots,6$;
$a_{\psi1}=0.01$, and $a_{\psi2}=0.01$. $S_j=\{\bbeta_j|\bw
_i^T\bbeta
_j>0, i=1,\ldots,n\}$ for $j=1,3,4,6$, and $S_2=S_5=R^7$. We also set
$\bLamb_{0r}=0$, $\bSig_{\Lambda r}=10^6$, $a_{1r}=3$, and $a_{2r}=1$
in (\ref{mesprior}).

\subsection{Results}

Four resolutions were used in the MR algorithm described in
Section~\ref{mrsec}. For each resolution in the MR algorithm, 2000
burn-in samples
were discarded and another 4000 MCMC samples were acquired for
estimation and inference.
The MR algorithm improves the efficiency of the MCMC algorithm by
dramatically reducing the autocorrelations of the\break MCMC samples of most
parameters (see S2 in supplementary material [\citet{supp}] for
further details).
In this study, the selection of tuning parameter $\alpha$ in RBBS does
not affect the sampling algorithm much because the drift function does
not dominate the diffusion function.
Estimated posterior means (Est), standard errors (SE) and their
quotient (Z) based on the finest resolution are shown in Table~\ref
{realest}. The Z values that pass the false positive rate threshold
[\citet{Benjamini1995}] $q=0.05$ are highlighted in bold font, while those
between $q=0.05$ and $q=0.10$ are highlighted in bold and italic font.

\begin{table}
\tabcolsep=0pt
\caption{The estimates (Est), standard errors (SE) and
Est/SE (Z) in the case study}\label{realest}
\begin{tabular*}{\textwidth}{@{\extracolsep{\fill}}lcd{3.3}d{2.3}d{2.3}d{2.3}d{2.3}d{2.3}d{2.3}@{}}
\hline
& &\multicolumn{1}{c}{\textbf{Intercept}} &\multicolumn{1}{c}{\textbf{Caffeine}}& \multicolumn{1}{c}{\textbf{Negative}} & \multicolumn{1}{c}{\textbf{Positive}}&
\multicolumn{1}{c}{\textbf{Gender}}& \multicolumn{1}{c}{\textbf{Race}} &\multicolumn{1}{c@{}}{\textbf{Age}}\\
\hline
Relative & Est & 2.568 & 0.606 & 0.026 & 0.214 & 0.458 & 0.242 &
-0.083 \\
surge & SE & 0.188 & 0.189 & 0.149 & 0.121 & 0.173 & 0.179 & 0.149 \\
rate $\beta_1^*$ & Z & 13.669 & \multicolumn{1}{c}{\phantom{0}\textbf{3.203}} & 0.176 & 1.773 &
\multicolumn{1}{c}{\phantom{0}\textbf{2.649}} & 1.349 & -0.557 \\ [3pt]
Daytime & Est & 0.377 & 0.025 & 0.106 & -0.014 & 0.206 & -0.163 &
0.110 \\
equilibrium & SE & 0.049 & 0.030 & 0.058 & 0.044 & 0.043 & 0.056 &
0.045 \\
$\beta_2^*$ & Z & 7.624 & 0.827 & 1.839 & -0.324 & \multicolumn{1}{c}{\phantom{0}\textbf{4.777}}
& \multicolumn{1}{c}{\textbf{$\bolds{-}$2.944}} &
\multicolumn{1}{c}{\phantom{0}\textit{\textbf{2.414}}} \\[3pt]
Daytime & Est & 1.621 & -0.203 & -0.196 & -0.181 & -0.238 & -0.034 &
0.157 \\
instability & SE & 0.142 & 0.082 & 0.126 & 0.126 & 0.135 & 0.159 &
0.130 \\
$\beta_3^*$ & Z & 11.445 & \multicolumn{1}{c}{\textit{\textbf{$\bolds{-}$2.480}}} & -1.555 & -1.438
& -1.758 & -0.212 & 1.212 \\[3pt]
Relative & Est & 2.737 & 0.326 & 0.002 & 0.236 & 0.269 & 0.632 & 0.184
\\
dipping & SE & 0.262 & 0.342 & 0.193 & 0.156 & 0.224 & 0.258 & 0.220 \\
rate $\beta_4^*$ & Z & 10.448 & 0.954 & 0.010 & 1.518 & 1.200 &
\multicolumn{1}{c}{\phantom{0}\textit
{\textbf{2.447}}} & 0.837 \\ [3pt]
Nighttime & Est & -1.206 & 0.168 & 0.067 & -0.013 & 0.298 & -0.109 &
0.147 \\
equilibrium & SE & 0.115 & 0.063 & 0.075 & 0.084 & 0.073 & 0.106 &
0.072 \\
$\beta_5^*$ & Z & -10.459 & \multicolumn{1}{c}{\phantom{0}\textbf{2.678}} &
0.886 & -0.159 & \multicolumn{1}{c}{\phantom{0}\textbf
{4.081}} & -1.024 & 2.027 \\ [3pt]
Nighttime & Est & 1.135 & 0.702 & 0.286 & -0.030 & 0.060 & -0.113 &
-0.001 \\
instability & SE & 0.214 & 0.477 & 0.182 & 0.101 & 0.134 & 0.146 &
0.133 \\
$\beta_6^*$ & Z & 5.299 & 1.473 & 1.573 & -0.295 & 0.448 & -0.772 &
-0.006 \\[3pt]
$\psi$ & Est & 2.992 & \multicolumn{1}{c}{SE} & 0.114 & & & & \\
\hline
\end{tabular*}
\end{table}

\begin{figure}

\includegraphics{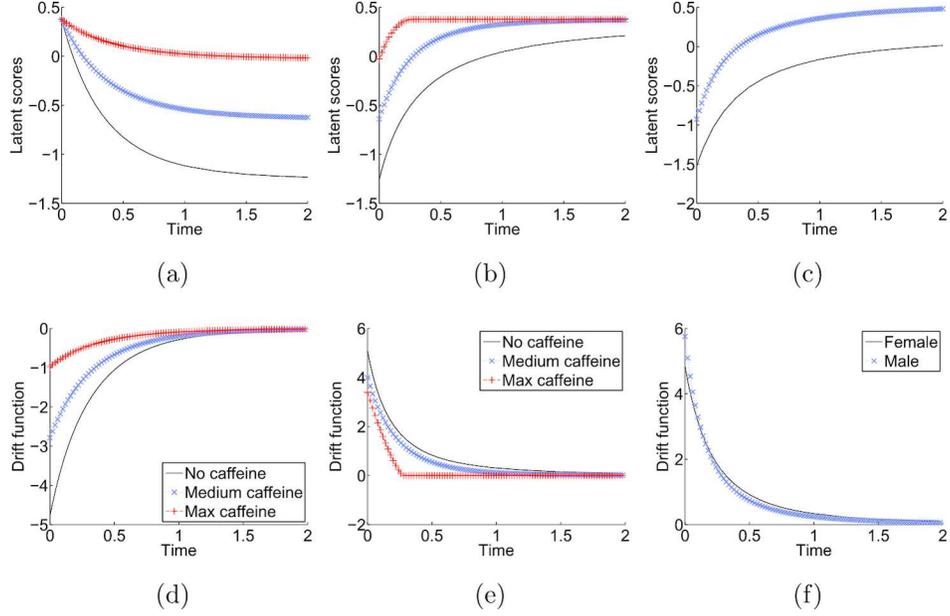}

\caption{The first and second rows are the estimated mean
trajectories and drift functions of CV trajectories, respectively. The
first and second columns show the night dipping and morning surge given
different levels of caffeine consumption, respectively. The third
column displays the gender differences during morning surge.}
\label{Caff-fig}
\end{figure}


Some covariates were found to be significantly correlated with one or
more aspects of the participants' CV trajectories. Based on the
estimated coefficients, we numerically simulated the mean trajectories
of subjects with different levels of certain covariates from (\ref
{dwm}), which are plotted in Figure~\ref{Caff-fig}.\ For continuous
covariates, three levels of covariates were used, namely, the minimum,
median and maximum of the covariates. Table~\ref{tab:estbphr} shows the
estimated equilibriums of SBP, DBP and HR at daytime and nighttime for
different levels of covariates.

\begin{table}[b]
\caption{Estimated equilibriums of SBP, DBP and HR at daytime and
nighttime for different levels of covariates. For continuous
covariates, the first and second rows are equilibriums at the minimum
and maximum of the covariates. For gender, the first row is female's
equilibriums. For race, the first row is black subjects' equilibriums}
\label{tab:estbphr}%
\begin{tabular*}{\textwidth}{@{\extracolsep{\fill}}ld{3.1}d{3.1}d{3.1}d{3.1}d{3.1}@{}}
\hline
& \multicolumn{2}{c}{\textbf{Night equilibrium}} & \multicolumn{3}{c@{}}{\textbf{Morning
equilibrium}} \\[-6pt]
& \multicolumn{2}{c}{\hrulefill} & \multicolumn{3}{c@{}}{\hrulefill} \\
& \multicolumn{1}{c}{\textbf{Caffeine}} & \multicolumn{1}{c}{\textbf{Gender}} & \multicolumn{1}{c}{\textbf{Gender}} & \multicolumn{1}{c}{\textbf{Race}} &
\multicolumn{1}{c@{}}{\textbf{Age}} \\
\hline
SBP & 100 & 95.2 & 125.1 & 131.9 & 126.2 \\
& 121.9 & 105.8 & 132.5 & 123.9 & 132.5 \\[3pt]
DBP & 58.7 & 55.2 & 77.4 & 82.4 & 78.1 \\
& 71.7 & 63.1 & 82.8 & 76.4 & 82.8 \\[3pt]
HR & 68 & 66.4 & 76.5 & 78.7 & 76.8 \\
& 75.4 & 70 & 78.9 & 76 & 78.9 \\
\hline
\end{tabular*}\vspace*{6pt}
\end{table}
%


We calculated the posterior predictive $p$-value [\citet
{gelmanposterior1996}] of the $\chi^2$ goodness-of-fit measure:
\[
D(\bY,\bX,\bLamb,\bSige)=\sum_{i=1}^n\sum
_{j=1}^{T_i}(\byij -\bLamb\bxij
)^T\bSige^{-1}(\byij-\bLamb\bxij).
\]
%
Let $\underline{\bX}_k$, $\underline{\bLamb}_k$ and $\underline
{\bSig
}_{\varepsilon k}$ be the $k$th MCMC sample of $\bX$, $\bLamb$ and
$\bSige
$, respectively. $\underline{\bY}_k$ was generated based on (\ref
{dwmm}) with $\underline{\bX}_k$, $\underline{\bLamb}_k$ and
$\underline
{\bSig}_{\varepsilon k}$. The estimated posterior predictive $p$-value is
the proportion of $D(\underline{\bY}_k,\underline{\bX}_k,\underline
{\bLamb}_k,\underline{\bSig}_{\varepsilon k})$ that are greater than
$D(\bY
,\underline{\bX}_k,\underline{\bLamb}_k,\underline{\bSig
}_{\varepsilon
k})$ for all MCMC samples (0.485), indicating a reasonable fit of the
model to the data. We also checked the predictive performance of our
model. The pointwise estimated median and 95\% credible\vadjust{\goodbreak} intervals
formed by all $\underline{\bY}_k$ for three randomly selected subjects
are displayed in Figure~S3. The prediction is reasonably
good even in the presence of sparsely spaced observations.

The acceptance rates of the cross-resolution move are 0.04, 0.14 and
0.27 at the 2nd, 3rd and 4th resolution, respectively. The increasing
acceptance rates agree with \citet{kou2011}. As $k$ increases, the
empirical distribution for $P_{k-1}(\bthetaa,\bX^{(k-1)}|\bY)$ becomes
a better proposal distribution because the difference between
$P_{k-1}(\bthetaa,\bX^{(k-1)}|\bY)$ and $P_k(\bthetaa,\bX
^{(k)}|\bY
)$ decreases.

The estimation is not very sensitive to the specification of the
hyperparameters in (\ref{mesprior}). Specifically, we set the
covariance matrices of the multivariate normal prior distributions to
$10\bI$, which represents moderate prior covariances, and changed the
prior mean to two times or half of the estimated parameters. The
estimation results were similar to Table~\ref{realest}.

%
%
%
%
%

\subsection{Substantive findings}

One interesting finding is that the estimated intercept of $\beta_3^*$
is significantly greater than 1. From a modeling aspect, this
contradicts the common assumption of the linear OU process and
demonstrates the added value of the nonlinear model considered.
Practically, compared to the dipping at night, the daytime surge is
much faster at the beginning and slower at the end, resulting in a less
stable change.

The covariates were found to play discrepant roles in affecting the
subjects' daytime and nighttime CV dynamics.
To further shed light on the substantive implications of such
differences, we compared two aspects of the modeling results thought to
be important from a CV standpoint: (i) daytime and nighttime
equilibrium levels of CV activities (modeled by $\beta_2^*$ and $\beta
_5^*$); (ii) relative rates and the instability of the surge and
dipping of CV activities. 

Caffeine and gender were related to the CV activities equilibrium at
night, while gender, race and age are related to the daytime equilibrium.
(i) The effects of caffeine intake in the literature remain mixed. For
instance, \citet{JOIM:JOIM683} reported that ``habitual coffee drinking
did not influence the 24-hour blood pressure profiles.'' However,
\citet
{Green1996cafraey} found that caffeine intake affected daytime SBP as
well as DBP, and nighttime BP. \citet{lane2002caffeine} showed
increased levels of ABP persisting for a few hours following caffeine
consumption. In this study, we found that caffeine significantly
increased the CV activities equilibrium at night ($\beta_{51}$), which
is shown in Figure~\ref{Caff-fig}(a). In contrast, caffeine intake in
the daytime did not affect the CV activities equilibrium in the morning
[Figure~\ref{Caff-fig}(b)].
(ii) \citet{Carels00a} showed that effect of negative emotion increases
the whole day CV activities. However, the daytime and nighttime CV
activities were not studied separately. In our study, the daytime
equilibrium was not related to negative emotion. (iii) We found that
male subjects have higher equilibrium in both day and night. Studies
using 24-hour ABP have shown
that BP is higher in men than in women at similar ages [\citet
{Reckelhoff01052001}]. (iv)~Elder subjects have higher daytime
equilibrium. In the US population, SBP increases progressively with
age, and DBP peaks at around age 55 years. Central arterial stiffening
with age is considered to account for this phenomenon [\citet
{Franklin01071997}]. (v) Minority subjects exhibit higher daytime
equilibrium. However, black and white subjects do not show much
difference in the nighttime equilibrium. \citet{Hinderliter01012004}
reported that African Americans have a smaller nocturnal decline in BP
than white subjects.

The disagreement between \citet{Hinderliter01012004} and our result
regarding race difference in nighttime equilibrium may be explained by
the relative rate of change found to be smaller for black subjects in
our model. In \citet{Hinderliter01012004}, the daytime and nighttime BP
were defined to be the average BP when subjects are awake or asleep. As
illustrated in Figure~\ref{rti-fig}(a), the average score is larger for
subjects with a smaller rate of change even when the daytime and
nighttime equilibriums are identical. In this aspect, our result agrees
with \citet{Hinderliter01012004}, which may shed light on the advantage
of our data analysis by considering the change of CV activities in
addition to equilibriums.

The rate of change of CV activities was seldom studied in the
literature. (i)~We found that white subjects have a faster nighttime
relative dipping rate. The pattern of latent scores and drift functions
are similar to those shown in Figure~\ref{rti-fig}(a) and~(b). (ii) The daytime surge rate and stabilization parameter are
affected by daytime caffeine intake. Although the absolute rates of
change of caffeine users are relatively small due to the smaller
difference between the two equilibriums, relative morning surge rates
are larger and the surge rates are more stable. The patterns of latent
scores and drift functions for subjects with different levels of
caffeine consumption are shown in Figure~\ref{Caff-fig}(b) and (e). (iii) Male subjects have a higher relative surge rate.
The latent scores and drift functions for male and female are shown in
Figure~\ref{Caff-fig}(c) and (f).

\section{Conclusion}

We have presented a latent SDE framework for population dynamic data
with covariates measured at irregularly spaced time points. Using the
Euler--Maruyama approximation to generate numerical solutions of the
SDEs, several local and global samplers for sampling the latent
processes have been developed based on modifications of existing
samplers in the literature.
The proposed model has been applied to the Duke Biobehavioral
Investigation of Hypertension study. Significant covariate effects have
been identified. Risk factors of Coronary heart disease (CHD) in
different scenarios have been discussed. 
Several scientific findings in our analysis were not evident from
previous studies. First, more information of the population dynamic
data is revealed in our study. Instead of studying the difference
between daytime and nighttime CV activities, we investigate several
dynamic characteristics of CV activities, including the daytime and
nighttime equilibriums, rates of change and change stability during the
night dipping and morning surge, respectively. Second, we explain the
variation among the population dynamic data with the effects of
covariates. We identify covariates that are correlated with these
dynamic characteristics. Third, the covariates effects for the
characteristics in daytime and nighttime are found to be asymmetric.

There are still some limitations of the proposed model, and further
developments are needed.
First, the processes between different subjects are assumed to be
independent in our model. One possible alternative is to consider
random effects variations of the proposed modeling framework to allow
for information borrowing among subjects.
Second, the time windows for capturing BP surges and dipping were
constructed based on subjective information. It would be more appealing
to develop data-driven methods to extract daytime and nighttime
windows, or to estimate the transition/change points empirically [e.g.,
\citet{Daniel1993Bayesian}].
Finally, the initial conditions of the SDEs were fixed to the estimated
latent scores for each subject at the first observed time point. Other
methods for approximating the unknown initial conditions [e.g.,
as mixed effects to be estimated as other modeling
parameters;
\citet{chowPsychometrika2015}] can, in principle, be used with
the proposed MR algorithm. The effects of using different approaches to
estimate the initial conditions of latent SDE models warrant further
investigation.

The MR algorithm can produce efficient MCMC chains as the resolution
increases. However, careful design and implementation are still
required for the MR algorithm to work properly and efficiently. The
proposal distribution consists of two parts, that is, the posterior
distribution at the previous resolution and the proposal distribution
of the processes at additional time points at the new resolution. The
posterior distribution depends recursively on the local updating
algorithm at the first resolution.
Thus, the performance of the local updating algorithms should be
reasonably satisfactory at the first resolution in order to provide
good building blocks for the MR algorithm. Otherwise, the empirical
distribution of the MCMC samples may not approximate the posterior
distribution well, and may not serve as a good proposal distribution in
the cross-resolution sampler, resulting in a low acceptance rate of the
cross resolution move and offsetting any potential advantages of the MR
algorithm. We use the sampler in \citet{Lindstrom2012} as the second
part, which contains a tuning parameter accounting for the nonlinearity
of the processes. We suggest using the same tuning parameter in the
local sampler with high acceptance rate.

The efficiency of the MR algorithm may be reduced when the number of
subjects is large. The MR algorithm is essentially a MH algorithm,
which updates the processes for all subjects at observed and imputed
time points and all parameters simultaneously. Maintaining the
acceptance ratio of a large number of random variables is challenging
for the MH algorithm. Algorithms to update the processes subject by
subject may be helpful.
In addition, other MCMC algorithms [e.g., \citet
{andrieu2010particle,golightly2011bayesian,stramer2011bayesian}] that
sample the stochastic processes globally and do not lead to a
convergence problem when the number of imputation increases are worthy
of further research.

Another caveat is that even though the block updating scheme can
overcome the dependence between the latent states, the dependence
between the latent states and the parameters in the diffusion function
remains, as discussed by \citet{Roberts01102001}.\hskip.2pt\footnote{We thank a
reviewer for pointing this out.} In our studies, the MCMC algorithms
work satisfactorily at the first resolution, and the MR algorithm keeps
improving the MCMC algorithms as the number of imputations increased.
However, when the 1-step and block updating algorithm fail at the first
resolution, other alternative MCMC algorithms should be considered as
the building block for the MR resolution to work properly.


\begin{supplement}[id=suppA]
\stitle{Supplementary materials of ``Bayesian analysis of ambulatory
blood pressure dynamics with application to irregularly spaced sparse data''}
\slink[doi]{10.1214/15-AOAS846SUPP} 
\sdatatype{.pdf}
\sfilename{aoas846\_supp.pdf}
\sdescription{We provide details of the MCMC algorithms and additional
analysis of the case study}
\end{supplement}

\section*{Acknowledgment}
Address for correspondence and reprints: Dr. Hongtu Zhu, e-mail: \printead*{e4}; phone: 919-966-7272.
%

\printaddresses
\end{document}